# Consistent Segregation Metrics: Addressing Structural Variations in Global Labor Markets

Ana Kujundzic[*] and Janneke Pieters[†]

**Abstract**

The Index of Dissimilarity (ID), widely utilized in economic literature as a measure of segregation, is inadequate for cross-country or time series studies due to its failure to account for structural variations across countries' labor markets or changes over time within a single country's labor market. Building on the works of Karmel and MacLachlan (1988) and Blackburn et al. (1993), we propose a new measure—the standardized ID—that isolates structural differences from true differences in segregation across space or time. A key advantage of our proposed measure lies in its ease of implementation and interpretation, even when working with datasets encompassing a large number of countries or time periods. Moreover, our measure can be consistently applied in the case of lumpy sectors or occupations that account for a large fraction of the workforce. We illustrate the new measure in an analysis of the cross-country relationship between economic development (as measured by GDP per capita) and occupational and sectoral gender segregation. Comparing the crude ID with the standardized ID, we show that the crude ID overestimates the positive correlation between income and segregation, especially between low- and middle-income countries. This suggests that analyses relying on the crude ID risk overestimating the importance of income differentials in explaining cross-country variation in gender segregation.

[*] Corresponding author. This manuscript was developed as part of the author's doctoral research at Wageningen University and is included as a chapter in the author's PhD thesis titled *Understanding Income and Labor Market Inequalities: Methods and Applications*. E-mail: akujundzic.1@gmail.com.

[†] CPB, Wageningen University, and IZA.

# 1 Introduction

Occupational segregation by gender is a key feature of labor markets across the world and an important determinant of gender wage inequality (Anker 1997; Blau and Kahn 2017; Borrowman and Klasen 2020). A recurring question in the literature is how economic development affects occupational segregation, an answer to which would help uncover the ultimate drivers of gendered segregation and develop policies to reduce gender inequality and labor allocation inefficiencies (Anker et al. 2003; Bandiera et al. 2022; Blau et al. 2013; Borrowman and Klasen 2020).

As noted by Blackburn et al. (1995) and Elbers (2023), however, measures of gender-based occupational segregation applied in the literature are not well comparable across countries, or across time periods within countries. For example, the most popular measure of gender-based occupational segregation, the Index of Dissimilarity (ID), also known as Duncan Index, is sensitive to changes in occupational structure and (except under specific conditions) to changes in female labor force participation.

Both occupational structure and female labor force participation vary widely across countries and within countries over time. Economic growth is typically accompanied by substantial changes in the occupational and sectoral structure of the economy, as the importance of agriculture declines and the variety of occupations increases, especially in early stages of economic development (Bandiera et al. 2022). Female labor force participation (FLFP) has a well-known U-shaped relationship with GDP per capita across countries (World Bank 2011), while the pace and timing of FLFP growth within countries is itself a widely studied topic (e.g., Bhalotra and Fernández 2024; Blau and Kahn 2007; Gaddis and Klasen 2014).

To understand the relationship between economic development and gender labor market segregation, it is critical to use a segregation measure that is consistent across countries and different points in time. This requires separating true changes in segregation from mechanical changes stemming from shifts in occupational structure and female labor force participation. Several solutions have been proposed so far, including decomposition methods by Gibbs (1965), Fuchs (1975), and Karmel and MacLachlan (1988), as well as a method based on so-called marginal matching (Blackburn et al. 1993). However, virtually all studies of occupational segregation in economics still rely on the crude ID measure without accounting for these mechanical changes.[1]

In this paper, building on the work of Karmel and MacLachlan (1988) and Blackburn et al. (1993), we develop a new method to identify true differences in occupational or sectoral gender segregation across space or time, which remains consistent even in the case of "lumpy" occupations or sectors. These are occupations or sectors that account for a large fraction of the workforce, such as agriculture in low-income countries, and create inconsistencies in previous marginal matching based measures. The primary advantage of our method lies in its simplicity of implementation and interpretation compared to other approaches reviewed in this study. This is particularly relevant for cross-country comparative studies of occupational or sectoral segregation, especially when working with a large number of countries.

We illustrate our method in an analysis of the cross-country relationship between economic development and gender labor market segregation, analysing cross-country patterns in occupational and sectoral gender segregation. We find that variation in the female share of the workforce and in the occupational or sectoral structure of the economy explains a substantial share of the cross-country variation in the gender labor market segregation as measured by the crude ID. This is especially true for the variation between low- and middle-income countries. As a result, our new standardized ID measure indicates a much weaker relationship between per capita GDP and gender labor market segregation compared to the crude ID measure.

Our paper not only contributes to the literature on economic development and gender segregation, but also to the literature on labor market segregation measurement by introducing a new method to analyze cross-country differences in occupational or sectoral gender segregation, as well as time trends in segregation within countries. This method can also be applied to study differences across space or time on other binary dimensions of segregation, such as race. Specifically, we propose a new solution to the well-known margin-dependency problem of the crude ID measure of segregation: its sensitivity to differences in occupational structure and female labor force participation. The basic idea behind our approach is to utilize Basic Segregation Tables to calculate the crude ID, as proposed by Blackburn et al. (1993), and then standardize them using the Iterative Proportional Fitting (IPF) procedure. Standardizing the basic tables enables us to isolate variation in the marginal component (occupational structure and female labor force participation) across space or time, so that the remaining differences in the crude ID can be attributed solely to differences in true segregation. As mentioned earlier, the primary advantage of our proposed method is its relative ease of implementation and interpretation, and its consistency in the case of lumpy sectors or occupations.

---

[1] An exception is Blau et al. (2013), who apply Fuchs' (1975) decomposition method to study trends in occupational segregation in the US from 1970 to 2009.

## 2 Measuring Occupational Gender Segregation

In this section, we review the literature on measures of labor market segregation, with a focus on the Index of Dissimilarity, also known as Duncan Index, the most widely used measure of segregation in the economics literature. We discuss the so-called invariance problem and review previous solutions proposed in the literature. We then develop a new and simple method to tackle the invariance problem, which ensures comparability across countries or over time even in the case of lumpy occupations. Throughout this section, we refer to occupational segregation by gender, but the methodology equally applies to sectoral segregation by gender and other (binary) dimensions of segregation such as ethnic background.

### 2.1 Index of Dissimilarity

The Index of Dissimilarity (ID), introduced by Duncan and Duncan (1955), is the most widely used measure for studying occupational segregation by gender, mainly because it is fairly straightforward to calculate and interpret (e.g., Beller 1985; King 1992; Baunach 2002; Blau et al. 2013; Cortes and Pan 2018; Bandiera et al. 2022). Subsequent to its introduction, numerous studies have scrutinized the properties of the ID and other measures of segregation, while also developing criteria to evaluate their efficacy (e.g., James and Taeuber 1985; Charles and Grusky 1995; Grusky and Charles 1998). Blackburn (2012) lists seven key criteria for the measurement of occupational gender segregation: Gender Symmetry (men and women should be equally segregated from each other), Constant Upper Limit (the upper value of the measurement range should be fixed at one or 100 percent), Constant Lower Limit (the lower value of the range should be fixed at zero), Size Invariance (the total number of workers should not affect the measured level of segregation), Occupational Equivalence (if two or more occupations have the same gender composition, it should not affect the measure of segregation whether these occupations are treated separately or combined in one), and finally the criteria of Sex Composition Invariance and Gendered Occupations Invariance. It is well known that the ID measure fails to satisfy both of these last two invariance criteria, which renders it problematic to use for comparing segregation levels across countries or over time within a country.

To better understand the invariance problem and some of the proposed solutions in the literature, as well as our own proposed solution, it is helpful to work with the Basic Segregation Table. This table was first introduced by Blackburn et al. (1993), who conceptualized segregation as the tendency for women to work in female-dominated occupations (or sectors) and men in male-dominated ones. The Basic Segregation Table, as shown in Table 1, is a (2x2) cross-tabulation of workers by occupation and sex. Occupations are grouped into two gendered categories—male and female occupations.

If we denote the number of women in occupation $i$ by $F_i$ and the total number of workers in occupation $i$ by $N_i$, female occupations are defined as those with a higher proportion of women than the proportion of women in the workforce ($F_i / N_i > F / N$). Similarly, male occupations are defined as those with a higher proportion of men than the proportion of men in the workforce ($M_i / N_i > M / N$). In other words, female occupations are those where women are overrepresented, and male occupations are those where men are overrepresented, relative to the total workforce.

Table 1. Basic Segregation Table: 2x2 cross-tabulation of workers by occupation and sex

|                     | Women | Men   | Total |
|---------------------|-------|-------|-------|
| Female Occupations  | $F_f$ | $M_f$ | $N_f$ |
| Male Occupations    | $F_m$ | $M_m$ | $N_m$ |
| Total               | F     | M     | N     |

*Note*: F is the number of female workers; M is the number of male workers; N is the total number of workers; subscript f denotes the category of female occupations; subscript m denotes the category of male occupations.

The ID measure can be thought of as a statistic of association between two variables of the Basic Segregation Table (gendered occupations and workers' sex)[2]—indicating the degree of concentration of women in female occupations and of men in male occupations. The stronger the relationship between gendered occupations and workers' sex, the higher the level of segregation.

Within the context of the Basic Segregation Table, ID is defined as the difference of proportion between the table columns.[3] In other words, it is the difference in the proportion of all women and men employed in female occupations. Or, equivalently, the difference in the proportion of all men and women employed in male occupations:

$$ID = \frac{F_f}{F} - \frac{M_f}{M} \equiv \frac{M_m}{M} - \frac{F_m}{F} \qquad (1)$$

The ID measure ranges from zero to one. A value of zero represents complete absence of segregation, indicating that the proportions of women and men employed in female (or male) occupations are equal. Conversely, a value of one signifies complete segregation, meaning that all women are in female occupations and all men are in male occupations.

---

[2] Or, in the case of sectoral gender segregation, gendered sectors and workers' sex.

[3] Blackburn et al. (1993, p. 360) show that this formulation of ID is mathematically equivalent to the conventional formula for ID $\left(ID = \frac{1}{2}\sum_i |F_i/F - M_i/M|\right)$.

## 2.2 Solution to the Invariance Problem

As previously mentioned, the ID measure fails to satisfy both the sex composition invariance and gendered occupations invariance criteria, making it problematic to use in cross-country or time series studies of segregation. According to Blackburn (2012), the sex composition criterion requires that the measure of segregation is not directly influenced by the overall gender composition of the workforce (column totals F and M in Table 1). Gendered occupations invariance requires that the measure of segregation is not directly influenced by the total number of workers in female and male occupations (row totals $N_f$ and $N_m$ in Table 1).

These two invariance criteria are important because they ensure that the measure of segregation is not affected by any linear transformations of the data. Consider a labor market with men and women distributed across two occupations at two points in time, as shown in panels A and B of Table 2.

Table 2: Linear transformation of the data example

|            | Panel A: year 1 | | | Panel B: year 2 | | |
|------------|---|---|---|---|---|---|
| Occupation | Women | Men | Total | Women | Men | Total |
| C | 10 | 70 | 80 | 15 | 70 | 85 |
| D | 60 | 30 | 90 | 360 | 120 | 480 |
| Total | 70 | 100 | 170 | 375 | 190 | 565 |

*Notes*: Occupation C in panels A and B is defined as a male occupation, while occupation D is defined as a female occupation in both panels. The odds ratio for panel A ($OR_A$) is calculated as: $OR_A = (10 \times 30)/(70 \times 60) = 0.07$. The odds ratio for panel B ($OR_B$) is calculated as: $OR_B = (15 \times 120)/(70 \times 360) = 0.07$. The index of dissimilarity for panel A ($ID_A$) is calculated as: $ID_A = (60/70) - (30/100) = 0.56$. The index of dissimilarity for panel B ($ID_B$) is calculated as: $ID_B = (360/375) - (120/190) = 0.33$.

The data in panel B is a linear transformation of the data in panel A, formed by column scaling (tripling the number of women and doubling the number of men) and row scaling (halving the size of occupation C and doubling the size of occupation D). The two panels are equivalent in that they have the same degree of association between occupation and gender. We can conceptualize panel A as containing a "basic nucleus" that describes its association, with all other data derived from it (through row and column multiplications) sharing this same nucleus. Indeed, the odds ratio—a measure of association that is invariant to row and column multiplications—is identical in both panels ($OR_A = OR_B = 0.07$).[4] In contrast, the ID measure varies ($ID_A = 0.56$; $ID_B = 0.33$) because it is sensitive to column scaling (i.e., changes in the gender composition of the workforce) and row scaling (i.e., changes in the size of occupations).

One solution to the invariance problem of ID that can be found in the literature is to decompose the overall difference in ID across time or space into marginal differences and segregation differences (Watts 2005). The marginal component quantifies differences in the gender composition of the workforce and the size of occupations. The segregation component quantifies differences in the true level of segregation. The basic idea behind decomposition is to isolate variation in the marginal component so that the remaining differences in ID can be attributed solely to differences in true segregation.

One of the earliest examples is a paper by Gibbs (1965), who uses the ID measure to study occupational segregation of white and black men across 50 states of the United States (US). Recognizing that differences in the size of occupations across states might conflate with true differences in segregation, he standardizes the size of occupations by assigning 1,000 workers to each occupational category, maintaining the same racial ratio in each category as observed in the census data. Another example is a paper by Fuchs (1975), who employs a similar standardization method to study changes in occupational gender segregation among "professional, technical, and kindred workers" in the US from 1950 to 1970. Instead of assigning 1,000 workers to each occupational category like Gibbs, Fuchs standardizes the size of occupations across two decades by fixing them at the 1960 values.

None of these methods are fully satisfactory because they do not take into consideration the sex composition criterion (or race composition criterion in Gibbs' example). In fact, it has often been mentioned that the ID measure meets the requirement of sex composition invariance, and therefore we need not worry about it (see James and Taeuber 1985, and Borrowman and Klasen 2020). However, this statement is true only if changes in the size of the female or male workforce leave the distribution of female and male workers across occupations unaffected. As Blackburn et al.

---

[4] The odds ratio is the ratio of the product of the main diagonal elements to the product of the minor diagonal elements of a (2x2) contingency table. Given the (2x2) table [a b; c d], the odds ratio is defined as $ad/bc$.

(1993, p. 347) put it, "such circumstances are so implausible that independence in this respect is virtually meaningless, and is certainly not an adequate definition of sex composition invariance."[5]

To meet both the sex composition invariance and gendered occupations invariance criteria, Blackburn et al. (1993) suggest a new approach, which they refer to as the Marginal Matching (MM) procedure. Their approach involves matching the marginals (i.e., the row-total and the column-total) of the Basic Segregation Table, hence the name "Marginal Matching." This is done by using a different definition of male and female occupations from that associated with the Basic Segregation Table: male and female occupations are defined in such a way that the row-total and column-total marginals of the Basic Segregation Table are the same. Whatever happens to the gender composition of the workforce (the column-total marginal), the classification of occupations is adjusted to ensure the row-total marginal matches the column-total marginal. More intuitively, female occupations are defined as those employing the highest proportion of women, such that together they contain the same number of workers as there are women in the workforce ($N_f = F$). The remaining occupations are classified as male occupations. In his 2012 follow-up paper, Blackburn refers to this new (2x2) table as the Modified Segregation Table where $N_f = F$, $N_m = M$, and $F_m = M_f$.

There are two methodological issues with the MM approach. Firstly, the MM procedure does not control for differences in the marginals across time or space. The matching of the marginals only ensures that the Modified Segregation Table is symmetrical, which results in several statistics of association having the same value as the ID measure.[6] The Gini coefficient, phi-coefficient, and Kendall's tau-b correlation coefficient have the same value as ID when applied to the Modified Segregation Table because of table symmetry. These statistics are jointly referred to as the MM measure of segregation, with the value ranging from zero (no segregation) to one (total segregation). Although it is a nice feature that the value of ID coincides with the value of the phi-coefficient,[7] especially if segregation is conceptualized as the strength of the relationship between two variables of the Basic Segregation Table, the MM method does not control for differences in the marginals over time or across countries. Jarman et al. (1999) and Blackburn (2012) recognize this point by stating that the MM measure needs to be standardized to obtain comparability across different data sets. This assessment will become clearer in the next section when we discuss our proposed solution to the invariance problem.

---

[5] See Blackburn et al. (1990) and Blackburn and Marsh (1991) for a more extended discussion on this issue.

[6] See Appendix A for the proof.

[7] Phi-coefficient is mathematically equivalent to Pearson's correlation coefficient for (2x2) contingency tables.

A second methodological issue with the MM approach is what we call the "lumpy occupations problem" where the two sets of marginals of the Basic Segregation Table are not properly matched. To illustrate this point, consider a simplified labor market with 100 women and 200 men distributed across four occupations, as shown in panel A of Table 3. The (4x4) table can be collapsed into a (2x2) Basic Segregation Table by using $F/N = 1/3$ as the cutoff point to divide four occupations into male and female occupations. In occupations 1, 2, and 3, female workers account for more than 1/3 of the total workers, so these are female occupations, while occupation 4 is a male occupation (see panel B of Table 3).

To make the two sets of marginals of the Basic Segregation Table equal, the MM approach uses a different cutoff point than $F/N = 1/3$ to divide occupations into male and female ones. In the Modified Segregation Table, female occupations are those with the highest proportion of women (i.e., the highest $F_i/M_i$ ratio), which together contain the same number of workers as there are women in the workforce. In our hypothetical example, this means we take the occupations with the higher proportion of women until we reach a total of 100 workers. The lumpy occupation problem is illustrated by the fact that occupations 1 and 2 together employ only 65 workers, while if we add occupation 3 we have a total of 165 workers. Occupation 3 is relatively female-intensive, but it is a lumpy occupation with a very large share of the total workforce. Suppose we choose to classify only occupations 1 and 2 as female occupations in our hypothetical example. We would then arrive at the Modified Segregation Table as the one shown in panel C of Table 3.

Notice how the two sets of marginals in the modified table are not equal. The matching of the marginals using the MM approach was not precise because most of the workforce is lumped in occupations 3 and 4. Precise matching using the MM approach only works if occupations are more evenly distributed in terms of their size. The issue of lumpy occupations is particularly relevant in developing countries where the majority of the workforce is concentrated in agriculture. It is problematic even with detailed occupational data at the 3-digit ISCO classification level, when most workers in low-income countries are categorized as "agricultural and fishery workers" (see Table C1 in Appendix C).

Table 3: Lumpy occupations problem example

| Occupation (i) | Women ($F_i$) | Men ($M_i$) | Total ($N_i$) |
|---|---|---|---|
| Panel A: (4x4) table | | | |
| 1 | 20 | 5 | 25 |
| 2 | 30 | 10 | 40 |
| 3 | 40 | 60 | 100 |
| 4 | 10 | 125 | 135 |
| Total | F = 100 | M = 200 | N = 300 |
| | | | |
| Panel B: (2x2) basic table | | | |
| Female occupations | 90 | 75 | 165 |
| Male occupations | 10 | 125 | 135 |
| Total | F = 100 | M = 200 | N = 300 |
| | | | |
| Panel C: (2x2) modified table | | | |
| Female occupations | 50 | 15 | 65 |
| Male occupations | 50 | 185 | 235 |
| Total | F = 100 | M = 200 | N = 300 |

*Note*: $F_i$ is the number of female workers in occupation *i*; $M_i$ is the number of male workers in occupation *i*; $N_i$ is the total number of workers in occupation *i*.

## 2.3 IPF Basic Segregation Table Standardization

We propose an alternative approach to address the invariance problem of the ID measure of segregation: standardizing the Basic Segregation Table using the Iterative Proportional Fitting (IPF) algorithm.[8] Unlike the MM approach, which focuses solely on how occupations are classified into male and female categories, our method prioritizes the association structure between occupations and gender. Instead of employing a different cutoff point to match the two sets of marginals of the Basic Segregation Table, as the MM approach does, the IPF method iteratively scales the row and column elements of the Basic Segregation Table until the desired marginals are obtained. Importantly, the association structure of the initial table, as measured by the odds ratio, is preserved throughout the iteration process.

---
[8] The detailed workings of the IPF algorithm are explained in Kujundzic (2024) and in Appendix D.

While the IPF method has been widely used in statistics and computer science for standardizing contingency tables with non-uniform marginals since Mosteller proposed it in 1968, its application in segregation studies has been more recent and fairly limited. We are not aware of any segregation studies in the economics literature that apply the ID measure to the IPF-standardized Basic Segregation Tables to analyze cross-country variation in occupational and sectoral gender segregation. The first instance of the IPF method being used in this context is Karmel and MacLachlan's (1988) study on Australian occupational gender segregation (Watts 2005; Elbers 2023). In this study, the IPF algorithm is used to match the size of occupations and the number of men and women in the workforce in an earlier period (period 1) to those in a later period (period 2). However, the authors do not utilize Basic Segregation Tables or the ID as a measure of segregation. A similar application of the IPF can be found in a recent paper by Elbers (2023), which examines changes in occupational gender segregation in the US between 1990 and 2016, as well as changes in multigroup racial segregation in Brooklyn, New York City, from 2000 to 2010. Like Karmel and MacLachlan, Elbers does not incorporate Basic Segregation Tables or the ID measure in his analysis.

With that said, our proposed approach combines insights from Blackburn et al. (1993) and Karmel and MacLachlan (1988). The primary advantage of working with the Basic Segregation Table is its provision of a simpler and more intuitively appealing way for studying properties of various segregation measures and criteria for effective measurement, particularly the two invariance criteria. Additional benefits include the relative ease of calculating segregation indices, simplified implementation of the standardization method, and shorter computation time required to run the IPF algorithm for (2x2) Basic Segregation Table, compared to higher-dimensional tables as in Karmel and MacLachlan (1988) and Elbers (2023). These advantages are particularly relevant for comparative studies of occupational gender segregation across multiple countries. Furthermore, by standardizing the Basic Segregation Table using the IPF algorithm, we avoid two pitfalls of the MM method. Namely, the IPF method ensures comparability across time or space, and circumvents the lumpy occupations problem.

The choice of target marginals is arbitrary, provided they are identical across different tables. In other words, when two (or more) Basic Segregation Tables have identical marginals, any differences in the overall segregation levels can be attributed solely to differences in true segregation. Conversely, if these tables have different marginals, differences in overall segregation levels result from a combination of two factors: differences in the gender composition of the workforce and the size of occupations (i.e., marginal differences), and differences in true segregation. We provide a proof of this concept in Appendix B, along with the following illustrative example for a better understanding of the proposed method.

Suppose we want to know if there is any difference in the level of occupational gender segregation between countries A and B. The labor market of both countries can be described with the Basic Segregation Tables shown in panels A and B of Table 4.

Table 4: IPF Basic Segregation Table standardization example

|  | Women | Men | Total |
|---|---|---|---|
| Panel A: country A basic table |  |  |  |
| Female occupations | 310 | 120 | 430 |
| Male occupations | 110 | 460 | 570 |
| Total | 420 | 580 | 1,000 |
|  |  |  |  |
| Panel B: country B basic table |  |  |  |
| Female occupations | 400 | 385 | 785 |
| Male occupations | 45 | 170 | 215 |
| Total | 445 | 555 | 1,000 |
|  |  |  |  |
| Panel C: country B standardized table |  |  |  |
| Female occupations | 260 | 170 | 430 |
| Male occupations | 160 | 410 | 570 |
| Total | 420 | 580 | 1,000 |

Using the basic tables, we can calculate the ID measure of segregation for country A $(ID_A = (310/420) - (120/580) = 0.53)$ and for country B $(ID_B = (400/445) - (385/555) = 0.21)$. The question is, how much of the overall difference in ID can be attributed to differences in true segregation?

To identify the segregation component of the overall difference in ID, we use the IPF algorithm to standardize country B's Basic Segregation Table so that it has the same row-total and column-total marginals as country A's table (see panel C of Table 4). The standardized measure of segregation for country B (denoted by $SID_B$) is a regular ID measure applied to country B's standardized table ($SID_B = (260/420) - (170/580) = 0.33$).

Decomposition of the overall difference in ID between the two countries is as follows:

$$\begin{aligned} ID_A - ID_B &= \underbrace{(ID_A - SID_B)}_{segregation\ component} + \underbrace{(SID_B - ID_B)}_{marginal\ component} \\ &= (0.53 - 0.33) + (0.33 - 0.21) \\ &= 0.20 + 0.12 \\ &= 0.32 \end{aligned} \qquad (2)$$

From this decomposition, we can conclude that 63% of the overall difference in ID is due to differences in true segregation, while 37% is due to marginal differences (i.e., cross-country differences in the gender composition of the workforce and the size of occupations).

A well-known issue with decomposition analysis is the path-dependency problem, where the results of the decomposition depend on the choice of the reference category (Fortin et al. 2011; Elbers 2023). In this case, the results depend on the chosen target marginals. In the example above, we use country A's marginals as targets to standardize country B's Basic Segregation Table. We could have also chosen to standardize country A's Basic Segregation Table using country B's marginals as targets. Or, we could have standardized both countries' basic tables using a target marginal such as (1/2, 1/2). In the former case, we find that 30% of the overall difference in ID is due to differences in true segregation, while 70% is due to marginal differences.

As both the level of SID and the difference in SID between countries (or points in time) are affected by the choice of target marginals, one solution proposed by Deutsch et al. (2009) is to run multiple scenarios using different target marginals and then take a simple average of the decomposition results.

# 3 Application: Economic Development and Gender Segregation

To illustrate our method and its implications for research on gender segregation, we analyze variation in occupational gender segregation and sectoral gender segregation across countries. This application sheds light on the relationship between economic development (as measured by GDP per capita) and occupational and sectoral gender segregation, which has been studied widely in the economics literature, but mostly based on the crude ID (Anker et al. 2003; Borrowman and Klasen 2020; Bandiera et al. 2022). Comparing the crude and standardized ID, we illustrate to what extent cross-country variation in segregation is accounted for by differences in marginals versus differences in true segregation.

We focus on the descriptive cross-sectional relationship between economic development and gender segregation. The main reason for this is that panel data with sufficiently detailed occupational groups for a large number of countries is not available.[9] We construct and standardize Basic Segregation Tables so that each element of the row-total and column-total marginals is 1/2. The decision to use (1/2, 1/2) as the target marginal is driven by the fact that there is no natural "baseline" or "reference" category for cross-country comparison. Following Deutsch et al. (2009), we run three additional scenarios, using (1) a poor country's marginals (Uganda), (2) a middle-income country's marginals (Bolivia), and (3) a rich country's marginals (Switzerland) as target marginals for IPF standardization.

## 3.1 Data

We use harmonized census microdata from the International Integrated Public Use Microdata Series (IPUMS-International)[10] to construct Basic Segregation Tables and calculate the crude ID and standardized ID for each country. We include all employed, working-age individuals (15-64 years old) who are not residing in group quarters (e.g., hospitals, prisons, etc.) and who are not members of the armed forces. Harmonized occupational codes, based on the three-digit ISCO-88 classification system, and sectoral codes that roughly correspond to the broad structure (or sections) of the International Standard Industrial Classification (ISIC) system, are obtained from IPUMS-International.[11] The microdata are merged at the country-year level with data on GDP per capita in constant PPP-adjusted USD from the Penn World Table (PWT) version 10.01 database (Feenstra et al. 2015).

---

[9] Borrowman and Klasen (2020) use a panel of developing countries from the World Bank's I2D2 dataset, but this includes only 1-digit occupations and sectors.

[10] We wish to acknowledge the national statistical offices of all participating countries that originally produced and provided the data to the IPUMS-International.

The analysis is restricted to the most recent year for which data on three-digit ISCO-88 occupational codes, or sectoral codes, is available. The final working sample for the occupational segregation analysis includes 39 countries (refer to Table C*1* in Appendix C), whereas the sample for the sectoral segregation analysis includes 84 countries (refer to Table C*2* in Appendix C).

## 3.2 Results

Before we turn to the comparison of the crude and standardized ID, and in order to aid its interpretation, we first present the cross-country variation in the marginals of the Basic Segregation Table and in income (log GDP per capita). Figure 1 shows the female share of the workforce (panel A for the occupational segregation sample of 39 countries and panel C for the sectoral segregation sample of 84 countries), the share of workers in female occupations (panel B) and the share of workers in female sectors (panel D); all scatter plots have log GDP per capita on the horizontal axis. Remember in the Basic Segregation Table, female occupations (or sectors) are defined as those in which the female share of workers is greater than the female share of workers in the total workforce.

---

[11] The data contain up to 125 unique specified occupations and up to 17 specified industries.

Figure 1: Economic development and the Basic Segregation Table marginals

1A: Female share of workforce (N=39)
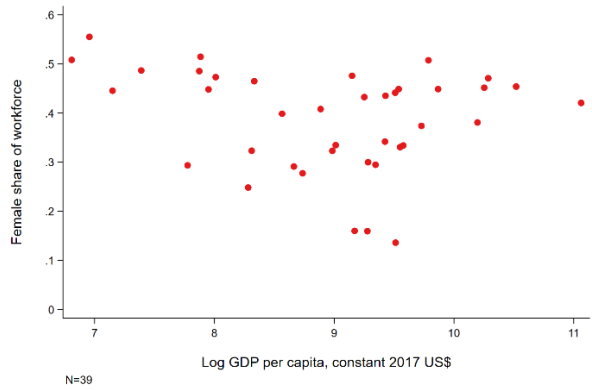

1B: Share of workforce in female occupations (N=39)
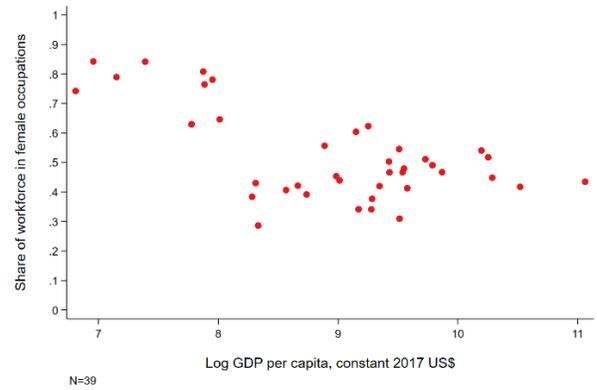

1C: Female share of workforce (N=84)
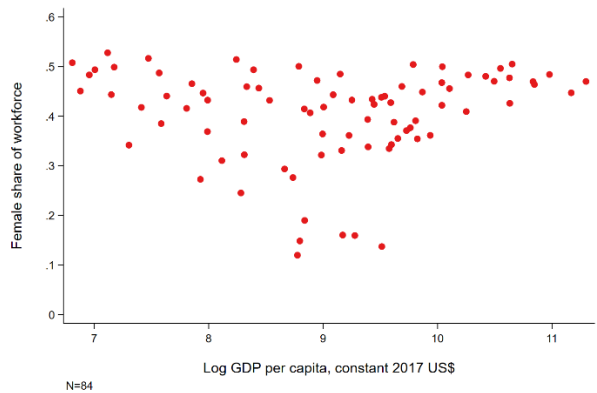

1D: Share of workforce in female sectors (N=84)
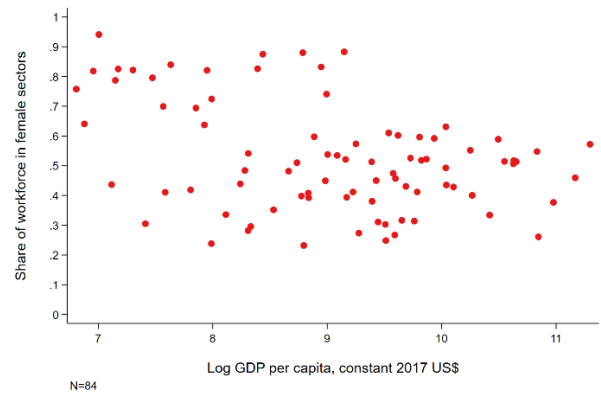

*Sources*: IPUMS-International, Penn World Table 10.01, and authors' calculations.

In panels A and C of Figure 1 we can see the well-known U-shape in female labor force participation across countries (see for example Goldin 1995; Gaddis and Klasen 2014; Dinkelman and Ngai 2021) as reflected in a lower female share of the workforce in middle-income countries. In both samples, women account for roughly 40 to 55 percent of the workforce in the very poor and very rich countries. In middle-income countries, the female share of the workforce ranges from 10 to 50 percent.

Panel B of Figure 1 shows a strong negative correlation between income and the share of workers in female occupations. We also find a negative correlation between income and the share of workers in female sectors, in panel D of Figure 1, but less strong. The high concentration of workers in female occupations and sectors in low-income countries is related to the importance of agriculture. Agriculture constitutes a high share of total employment in low-income countries, and the female share of workers in agriculture (and agricultural occupations) is typically higher than the female share of the total workforce (World Bank 2011).[12]

As economies grow from low- to middle-income level, we tend to see a reduction in the female share of the workforce and a decline in the relative size of female occupations. As a result, both the row and the column marginals of the Basic Segregation Table are quite different between low- and middle-income countries. From middle- to high-income, the female share of the workforce increases, while there is no clear trend in the relative size of female occupations and sectors.

How does variation in the marginals affect measured segregation? We now turn to the comparison of the crude and standardized ID. We start with occupational segregation. Figure 2 plots the relationship between income and occupational gender segregation across 39 countries with available data, using the crude ID and the standardized ID (with (1/2, 1/2) as the reference marginals). At low to middle income levels, there is a strong positive correlation between income and occupational gender segregation. But the crude ID overestimates the positive correlation. That is, the crude ID measures a greater increase in occupational gender segregation from low to middle income levels than the standardized ID. For middle to higher income levels, the correlation between income and occupational gender segregation is weak, and standardization has a limited effect on the cross country variation in segregation.

---

[12] This is closely related to the fact that a large share of agricultural work in low-income countries is home-based family farming (see Boserup 1970; Goldin 1995).

Figure 2: Occupational gender segregation and economic development

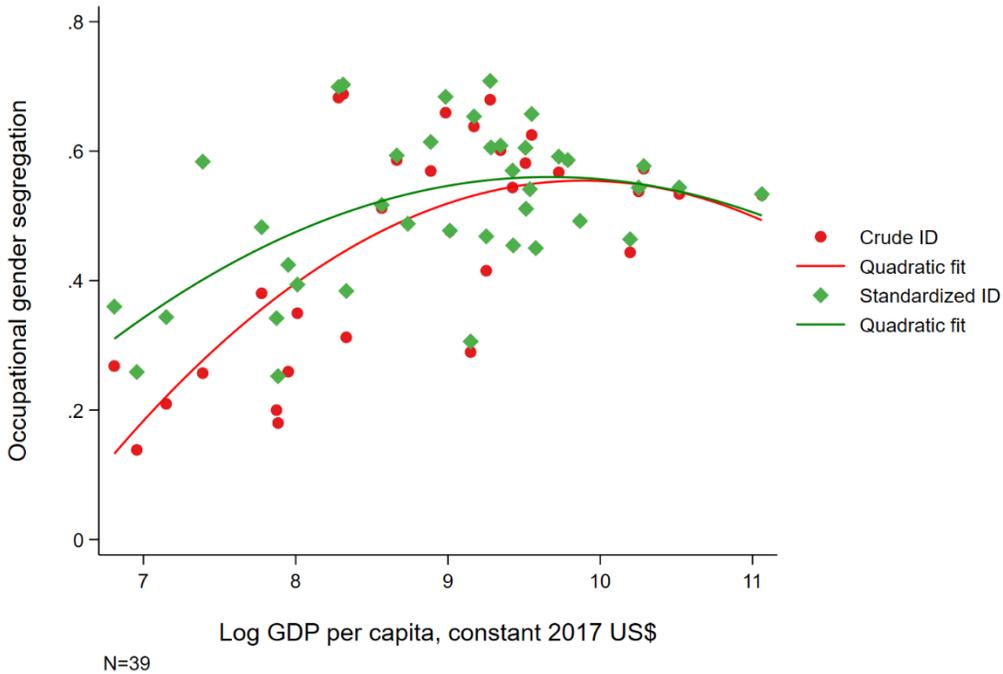

*Notes:* Crude ID is the ID measure of segregation calculated using Basic Segregation Tables. Standardized ID is the ID measure calculated using standardized tables with identical marginals (1/2, 1/2). *Sources*: IPUMS-International, Penn World Table 10.01, and authors' calculations.

A similar picture emerges for sectoral segregation (Figure 3), which we can analyze in a larger sample of 84 countries. Both the crude and the standardized ID have an inverse-U shaped relation with income. Variation in true segregation between low and middle-income countries (and among middle-income countries) is more limited than variation in the crude ID. The increase of the crude ID from low- to middle-income countries is thus partly driven by changes in the female share of workers and the relative size of female sectors. From middle-income to high-income the relationship between income and sectoral segregation is negative and looks slightly stronger with the standardized ID.

Figure 3: Sectoral gender segregation and economic development

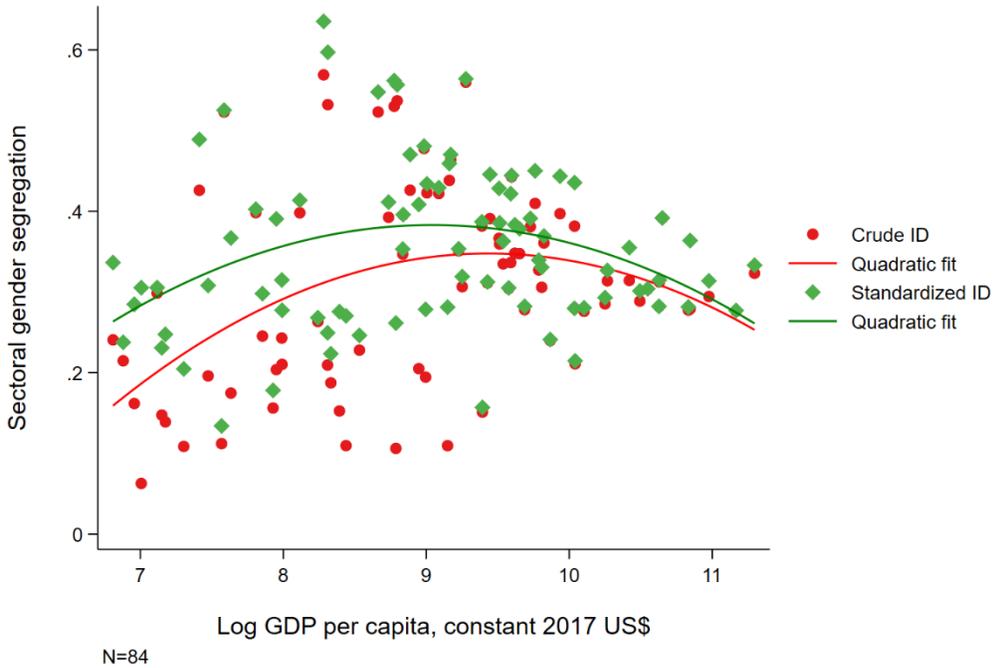

*Notes:* Crude ID is the ID measure of segregation calculated using Basic Segregation Tables. Standardized ID is the ID measure calculated using standardized tables with identical marginals (1/2, 1/2). *Sources*: IPUMS-International, PWT, and authors' calculations.

To analyze the relationship between GDP per capita and segregation more formally, and in order to assess sensitivity with respect to the choice of target marginals, we run a number of regressions. In panel A of Table 5, we analyze occupational gender segregation and regress both the crude ID (column 1) and standardized ID with target marginals 1/2 (column 2) on log GDP per capita and its square. In columns 3–5 the dependent variable is the standardized ID with different target marginals: those of Uganda in column 3 (F = 0.45, M = 0.55; $N_f$ = 0.79, $N_m$ = 0.21), those of Bolivia in column 4 (F = 0.40, M = 0.60; $N_f$ = 0.41, $N_m$ = 0.59), and those of Switzerland in column 5 (F = 0.42, M = 0.58; $N_f$ = 0.43, $N_m$ = 0.57).

We find that the relationship between log GDP per capita and its square with occupational segregation is weaker once the ID is standardized. Comparing column (1) to columns (2) to (5), both coefficients keep the same sign but they are closer to zero when the ID is standardized. The coefficients are smallest in column (3), where the ID is standardized using Uganda's marginals.

But regardless of the target marginal, we find that there is a substantially weaker relationship between per capita GDP and the standardized measure of occupational gender segregation, with the R-squared dropping from 0.516 with the crude ID to between 0.327 and 0.285 with the standardized ID.

Panel B of Table 5 reports regression estimates for sectoral gender segregation, again with the crude ID and standardized ID with different target marginals as dependent variables across the columns. Here, too, we see a decline in the size of the coefficients, which is most pronounced in column (3) when we use Uganda's marginals in the standardized ID. We also see a decline in the R-squared.

Table 5: Gender segregation and GDP per capita (pc)

|  | CID (1) | SID (1/2) (2) | SID (UG) (3) | SID (BO) (4) | SID (CH) (5) |
|---|---|---|---|---|---|
| Panel A: occupational gender segregation and GDP pc | | | | | |
| Log GDP pc | 0.880*** | 0.592** | 0.242** | 0.596** | 0.595** |
|  | (0.210) | (0.179) | (0.077) | (0.180) | (0.180) |
| Log GDP pc-squared | − 0.045*** | − 0.031** | − 0.012** | − 0.031** | − 0.031** |
|  | (0.012) | (0.010) | (0.004) | (0.010) | (0.010) |
| Constant | − 3.797*** | − 2.301** | − 0.880* | − 2.322** | − 2.315** |
|  | (0.922) | (0.790) | (0.343) | (0.794) | (0.793) |
| $R$-Squared | 0.516 | 0.286 | 0.327 | 0.285 | 0.285 |
| Observations | 39 | 39 | 39 | 39 | 39 |
| Panel B: sectoral gender segregation and GDP pc | | | | | |
| Log GDP pc | 0.516*** | 0.434*** | 0.216*** | 0.432*** | 0.434*** |
|  | (0.130) | (0.106) | (0.054) | (0.105) | (0.106) |
| Log GDP pc-squared | − 0.027*** | − 0.024*** | − 0.012*** | − 0.024*** | − 0.024*** |
|  | (0.007) | (0.006) | (0.003) | (0.006) | (0.006) |
| Constant | − 2.086*** | − 1.578** | − 0.753** | − 1.575** | − 1.580** |
|  | (0.581) | (0.468) | (0.240) | (0.465) | (0.468) |
| $R$-Squared | 0.166 | 0.111 | 0.110 | 0.111 | 0.111 |
| Observations | 84 | 84 | 84 | 84 | 84 |

*Notes*: Cross-country regressions with robust standard errors in parentheses. Column (1) reports the crude ID measure (CID). Columns (2) to (5) report the standardized ID measure (SID) with different target marginals: (1/2, 1/2) as the target marginal in column (2), Uganda's marginals in column (3), Bolivia's marginals in column (4), and Switzerland's marginals in column (5). Statistical significance levels: *$p < 0.05$; **$p < 0.01$; ***$p < 0.001$.

Our results show that for the crude ID measure of occupational and sectoral gender segregation, the cross-country variation that can be explained by log GDP per capita and its square is to quite a large extent accounted for by variation in the marginals of the Basic Segregation Table. The relationship between per capita GDP and the standardized ID is still significant and it has the same inverse-U shape, but the relationship is substantially weaker. This also implies that factors other than income level play a bigger role in explaining cross-country variation in gender segregation.

# 4 Conclusion

Margin dependency is a well-known problem in the study of differences in segregation across time and space. Even though several solutions have been proposed in the earlier literature, economists almost always rely on segregation measures that fail to satisfy Group Composition Invariance and Organizational Unit Invariance criteria.[13] This includes the most popular measure of segregation used by economists, the Index of Dissimilarity (ID), also known as the Duncan Index. The ID is a popular measure because it is straightforward to calculate and its interpretation is intuitive. However, differences in the ID across countries or time conflate differences in true segregation with differences in marginals. In the case of gender occupational segregation, this means that variation in the ID partly captures variation in the female share of workers and differences in the occupational structure of the economy.

In this paper, we propose a new measure, which we refer to as the standardized ID, that isolates the margin-independent component of the ID. Our method builds on the work of Karmel and MacLachlan (1988) and Blackburn et al. (1993). It is easy to implement, even when working with data for a large number of countries or time periods. Another advantage of the measure is that it can be consistently applied in the case of lumpy sectors or occupations. These are sectors or occupations that account for a large fraction of the workforce.

We illustrate the new measure in an analysis of the cross-country relationship between GDP per capita and occupational and sectoral gender segregation. Comparing the crude ID with the standardized ID, we show that the crude ID overestimates the positive correlation between income and segregation, especially between low- and middle-income countries. Regression analysis shows an inverse-U shaped relationship between log GDP per capita and segregation, which is much weaker for the standardized ID than for the crude ID. This holds for occupational segregation based on very detailed occupational groups as well as sectoral segregation based on a relatively coarse sector classification.

---

[13] Population groups could be differentiated by gender, race, age, and other characteristics. Organizational units could be sectors, occupations, firms, income ranges, etc.

Our findings imply that analyses of cross-country variation that use the crude ID, which has been a popular measure of gender segregation in the economics literature, risk overestimating the importance of differences in per capita income. The same is likely to hold for other country characteristics that have been associated with gender labor market segregation, such as the female labor force participation rate, gender gaps in education, or trade openness. To understand the extent to which such factors can account for variation in gender segregation, differences in true segregation need to be isolated from differences in the female share of the workforce and in the occupational or sectoral composition of the workforce. We expect that analysis of cross-country differences and time trends in gender segregation will remain important in future research on gender inequality, and we hope that the relatively simple method we propose will be adopted by scholars in this field.

# Appendix

## A. Proof: $ID = \Phi$ if (2x2) contingency table is symmetrical

Consider a (2x2) contingency table $X = [a\ b;\ c\ d]$.

The Index of Dissimilarity (ID) is defined as:

$$ID = \frac{a}{a+c} - \frac{b}{b+d} \tag{A.1}$$

The phi-coefficient is defined as:

$$\phi = \frac{ad - bc}{\sqrt{(a+b)(c+d)(a+c)(b+d)}} \tag{A.2}$$

If $X$ is symmetrical ($a = d;\ b = c$), then:

$$ID = \frac{a}{a+b} - \frac{b}{a+b} = \frac{a-b}{a+b} \tag{A.3}$$

and

$$\phi = \frac{a^2 - b^2}{\sqrt{(a+b)^4}} = \frac{(a-b)(a+b)}{(a+b)^2} = \frac{a-b}{a+b} \tag{A.4}$$

Hence, $ID = \phi$ if $X$ is symmetrical.

## B. Decomposition of differences in overall segregation

Consider a pair of (2x2) contingency tables $X = [a\ b; c\ d]$ and $Y = [p\ q; r\ s]$. We can compute the difference in the overall segregation level between the two tables by computing the difference between their ID measures. By expressing the ID measure in its logarithmic form for both tables and taking the difference of the logarithms, we obtain the following expression:

$$\begin{aligned}
\ln(ID_X) - \ln(ID_Y) &= \ln(a) - \ln(a+c) - \ln(b) + \ln(b+d) \\
&\quad - \ln(p) + \ln(p+r) + \ln(q) - \ln(q+s) \\
&= \underbrace{[\ln(a) - \ln(b) - \ln(p) + \ln(q)]}_{\Delta_1} \\
&\quad + \underbrace{[\ln(p+r) - \ln(q+s) - \ln(a+c) + \ln(b+d)]}_{\Delta_2}
\end{aligned} \quad (B.1)$$

We showed that the overall difference in the ID measure can be decomposed into two components, $\Delta_1$ and $\Delta_2$. The term $\Delta_1$ represents the segregation component, while $\Delta_2$ represents the marginal component. When the marginals of both tables are identical, $\Delta_2$ becomes zero, indicating that the difference in ID is purely due to segregation differences ($\Delta_1$). Conversely, if the marginals are not identical, then the overall difference in ID is due to both segregation and marginal differences.

## C. Datasets used in the analysis

Table C1: Countries and years included in occupational gender segregation analysis

| Country | Year | N | Largest occupation (% of workforce) |
|---|---|---|---|
| Armenia | 2011 | 102,837 | Market-oriented crop and animal producers (16%) |
| Belarus | 2009 | 422,310 | Motor-vehicle drivers (6%) |
| Bolivia | 2001 | 259,407 | Subsistence agricultural and fishery workers (25%) |
| Botswana | 2011 | 60,096 | Domestic and related helpers, cleaners and launderers (8%) |
| Cambodia | 2008 | 649,694 | Subsistence agricultural and fishery workers (63%) |
| Chile | 1992 | 430,105 | Market gardeners and crop growers (10%) |
| Costa Rica | 2000 | 126,196 | Agricultural, fishery and related laborers (11%) |
| Ecuador | 2001 | 409,247 | Agricultural, fishery and related laborers (13%) |
| Egypt | 2006 | 1,953,230 | Market gardeners and crop growers (23%) |
| El Salvador | 2007 | 181,701 | Shop salespersons and demonstrators (14%) |
| France | 1999 | 1,133,632 | Personal care and related workers (6%) |
| Greece | 2001 | 390,582 | Skilled agricultural and fishery workers (9%) |
| Guatemala | 2002 | 302,512 | Agricultural, fishery and related laborers (28%) |
| Guinea | 1996 | 257,329 | Subsistence agricultural and fishery workers (63%) |
| Honduras | 2001 | 161,189 | Market gardeners and crop growers (33%) |
| Iran | 2011 | 378,030 | Market gardeners and crop growers (12%) |
| Jordan | 2004 | 104,677 | Motor-vehicle drivers (9%) |
| Malaysia | 2000 | 158,626 | Market-oriented crop and animal producers (8%) |
| Mauritius | 2000 | 51,525 | Textile, fur, and leather products machine operators (10%) |
| Mongolia | 2000 | 77,073 | Subsistence agricultural and fishery workers (46%) |
| Mozambique | 2007 | 690,968 | Subsistence agricultural and fishery workers (73%) |
| Nicaragua | 2005 | 162,126 | Market gardeners and crop growers (19%) |
| Panama | 2000 | 107,748 | Market gardeners and crop growers (14%) |
| Paraguay | 2002 | 178,344 | Market gardeners and crop growers (22%) |
| Philippines | 2010 | 3,000,505 | Skilled agricultural and fishery workers (17%) |
| Portugal | 2001 | 238,723 | Numerical clerks (8%) |
| Romania | 2011 | 826,770 | Market gardeners and crop growers (7%) |
| Rwanda | 2002 | 315,151 | Subsistence agricultural and fishery workers (83%) |
| Senegal | 2002 | 265,709 | Market gardeners and crop growers (35%) |
| South Africa | 2007 | 195,672 | Domestic and related helpers, cleaners and launderers (13%) |
| Switzerland | 2000 | 135,377 | Secretaries and keyboard-operating clerks (6%) |
| Thailand | 2000 | 214,477 | Market gardeners and crop growers (21%) |
| Trinidad and Tobago | 2000 | 44,414 | Service workers and shop and market sales workers (9%) |
| Uganda | 2002 | 682,336 | Subsistence agricultural and fishery workers (69%) |
| United Kingdom | 1991 | 290,400 | General managers (7%) |
| Uruguay | 2006 | 101,958 | Domestic and related helpers, cleaners and launderers (9%) |
| Vietnam | 1999 | 1,059,523 | Agricultural, fishery and related laborers (64%) |
| Zambia | 2010 | 342,659 | Subsistence agricultural and fishery workers (34%) |
| Zimbabwe | 2012 | 203,369 | Subsistence agricultural and fishery workers (42%) |

*Notes*: N represents the final working sample size. Person-level sample weights are applied in the analysis.

Table C2: Countries and years included in sectoral gender segregation analysis

| Country | Year | N | Largest sector (% of workforce) |
|---|---|---|---|
| Argentina | 2001 | 1,008,551 | Wholesale and retail trade (18%) |
| Armenia | 2011 | 102,683 | Agriculture, fishing, and forestry (35%) |
| Austria | 2011 | 414,946 | Business services and real estate (16%) |
| Belarus | 2009 | 425,948 | Manufacturing (23%) |
| Benin | 2013 | 281,180 | Agriculture, fishing, and forestry (41%) |
| Bolivia | 2012 | 388,727 | Agriculture, fishing, and forestry (27%) |
| Botswana | 2011 | 61,281 | Public administration and defense (18%) |
| Brazil | 2010 | 8,242,200 | Wholesale and retail trade (18%) |
| Burkina Faso | 1996 | 394,098 | Agriculture, fishing, and forestry (89%) |
| Cambodia | 2019 | 792,360 | Agriculture, fishing, and forestry (54%) |
| Cameroon | 2005 | 439,668 | Agriculture, fishing, and forestry (58%) |
| Chile | 2017 | 642,244 | Wholesale and retail trade (18%) |
| China | 2000 | 6,466,684 | Agriculture, fishing, and forestry(63%) |
| Colombia | 2005 | 528,656 | Agriculture, fishing, and forestry (28%) |
| Costa Rica | 2011 | 162,632 | Wholesale and retail trade (19%) |
| Dominican Republic | 2010 | 386,888 | Wholesale and retail trade (24%) |
| Ecuador | 2010 | 505,176 | Agriculture, fishing, and forestry (22%) |
| Egypt | 2006 | 1,943,574 | Agriculture, fishing, and forestry (26%) |
| El Salvador | 2007 | 182,257 | Wholesale and retail trade (23%) |
| Fiji | 2014 | 32,940 | Agriculture, fishing, and forestry (40%) |
| Finland | 2010 | 112,469 | Health and social work (16%) |
| France | 2011 | 8,384,511 | Health and social work (14%) |
| Ghana | 2010 | 988,804 | Agriculture, fishing, and forestry (40%) |
| Greece | 2011 | 420,788 | Wholesale and retail trade (18%) |
| Guatemala | 2002 | 301,366 | Agriculture, fishing, and forestry (39%) |
| Guinea | 2014 | 323,563 | Agriculture, fishing, and forestry (51%) |
| Haiti | 2003 | 175,642 | Agriculture, fishing, and forestry (48%) |
| Honduras | 2001 | 159,236 | Agriculture, fishing, and forestry (40%) |
| Hungary | 2011 | 221,151 | Manufacturing (20%) |
| Indonesia | 2010 | 9,928,208 | Agriculture, fishing, and forestry (39%) |
| Iran | 2011 | 363,169 | Agriculture, fishing, and forestry (18%) |
| Iraq | 1997 | 380,795 | Other services (24%) |
| Ireland | 2016 | 192,884 | Wholesale and retail trade (15%) |
| Israel | 2008 | 200,201 | Manufacturing (15%) |
| Italy | 2011 | 1,218,224 | Manufacturing (16%) |
| Ivory Coast | 1998 | 500,540 | Agriculture, fishing, and forestry (56%) |
| Jamaica | 2001 | 64,210 | Wholesale and retail trade (15%) |
| Jordan | 2004 | 104,473 | Wholesale and retail trade (15%) |
| Kyrgyz Republic | 2009 | 220,864 | Agriculture, fishing, and forestry (45%) |
| Laos | 2015 | 313,931 | Agriculture, fishing, and forestry (75%) |
| Lesotho | 2006 | 47,005 | Agriculture, fishing, and forestry (32%) |
| Liberia | 2008 | 86,589 | Agriculture, fishing, and forestry (52%) |

| Country | Year | N | Largest sector (% of workforce) |
|---|---|---|---|
| Malawi | 2008 | 358,342 | Agriculture, fishing, and forestry (62%) |
| Malaysia | 2000 | 153,661 | Manufacturing (22%) |
| Mali | 2009 | 408,011 | Agriculture, fishing, and forestry (67%) |
| Mauritius | 2011 | 61,132 | Manufacturing (21%) |
| Mexico | 2015 | 3,457,086 | Wholesale and retail trade (18%) |
| Mongolia | 2000 | 78,759 | Agriculture, fishing, and forestry (47%) |
| Morocco | 2014 | 963,030 | Agriculture, fishing, and forestry (22%) |
| Mozambique | 2007 | 692,241 | Agriculture, fishing, and forestry (74%) |
| Myanmar | 2014 | 1,880,305 | Agriculture, fishing, and forestry (55%) |
| Nepal | 2011 | 1,200,222 | Agriculture, fishing, and forestry (66%) |
| Netherlands | 2011 | 229,912 | Other services (20%) |
| Nicaragua | 2005 | 161,691 | Agriculture, fishing, and forestry (33%) |
| Palestine | 2017 | 92,375 | Construction (21%) |
| Panama | 2010 | 132,498 | Wholesale and retail trade (20%) |
| Paraguay | 2002 | 176,030 | Agriculture, fishing, and forestry (26%) |
| Peru | 2017 | 1,157,897 | Wholesale and retail trade (19%) |
| Philippines | 2010 | 3,028,978 | Agriculture, fishing, and forestry (31%) |
| Poland | 2002 | 1,284,788 | Manufacturing (19%) |
| Portugal | 2011 | 243,178 | Wholesale and retail trade (17%) |
| Romania | 2011 | 826,770 | Agriculture, fishing, and forestry (22%) |
| Rwanda | 2012 | 400,626 | Agriculture, fishing, and forestry (76%) |
| Saint Lucia | 1991 | 5,381 | Agriculture, fishing, and forestry (21%) |
| Senegal | 2013 | 257,194 | Other services (51%) |
| Sierra Leone | 2004 | 173,165 | Agriculture, fishing, and forestry (66%) |
| Slovak Republic | 2011 | 305,450 | Manufacturing (26%) |
| Slovenia | 2002 | 70,639 | Manufacturing (31%) |
| South Africa | 2007 | 195,986 | Manufacturing (17%) |
| Spain | 2011 | 1,959,583 | Wholesale and retail trade (14%) |
| Sudan | 2008 | 1,406,590 | Agriculture, fishing, and forestry (33%) |
| Suriname | 2012 | 15,680 | Public administration and defense (19%) |
| Switzerland | 2011 | 125,318 | Manufacturing (15%) |
| Tanzania | 2012 | 1,646,407 | Agriculture, fishing, and forestry (62%) |
| Thailand | 2000 | 319,215 | Agriculture, fishing, and forestry (56%) |
| Togo | 2010 | 194,859 | Agriculture, fishing, and forestry (41%) |
| Trinidad and Tobago | 2000 | 43,909 | Wholesale and retail trade (13%) |
| Turkey | 2000 | 1,251,281 | Agriculture, fishing, and forestry (44%) |
| Uganda | 2002 | 685,305 | Agriculture, fishing, and forestry (74%) |
| United Kingdom | 2001 | 1,083,220 | Wholesale and retail trade (17%) |
| United States | 2015 | 8,194,483 | Wholesale and retail trade (14%) |
| Uruguay | 2006 | 102,807 | Wholesale and retail trade (19%) |
| Vietnam | 2019 | 4,294,657 | Agriculture, fishing, and forestry (34%) |
| Zambia | 2010 | 339,715 | Agriculture, fishing, and forestry (65%) |

*Notes*: N represents the final working sample size. Person-level sample weights are applied in the analysis.

# D. Iterative Proportional Fitting (IPF) algorithm

The following explanation of the IPF algorithm is taken from Kujundzic (2024).

Computationally, the IPF algorithm adjusts the cell frequencies (or relative frequencies) of the initial contingency table until the desired (i.e., target) marginal totals are achieved, while preserving the core pattern of association of the initial table, as measured by local odds ratios. An example of IPF standardization is illustrated in Figure D1, where the initial (2x2) relative contingency table is standardized so that its marginals match the target values of $f_{1+} = f_{2+} = f_{+1} = f_{+2} = 1/2$.

The first iteration step is a row scaling operation,[14] where each row cell of the initial table is multiplied by a scaling factor chosen in such a way that the sum of all row cells matches its target total:

$$f_{ij}^{(k)} = f_{ij}^{(k-1)} \times \frac{f_{i+}^*}{f_{i+}^{(k-1)}} \tag{D.1}$$

where $f_{ij}^{(k)}$ is the $i$th row cell of the $k$th iteration table, $f_{ij}^{(k-1)}$ is the $i$th row cell of the $(k-1)$ table, $f_{i+}^*$ is the $i$th row target total (1/2 in this example), and $f_{i+}^{(k-1)}$ is the $i$th row total of the $(k-1)$ table. After the first iteration ($k = 1$), the row totals are $f_{i+} = (1/2, 1/2)$, which matches the desired target.

The second iteration step is a column scaling operation, where each column cell of the first iteration table is multiplied by a scaling factor chosen so that the sum of all column cells matches its target total:

$$f_{ij}^{(k+1)} = f_{ij}^{(k)} \times \frac{f_{+j}^*}{f_{+j}^{(k)}} \tag{D.2}$$

---

[14] The choice of whether to start with the row scaling or the column scaling operation does not affect the final result of the IPF algorithm. The process will converge to the same standardized table regardless of the initial scaling operation.

where $f_{ij}^{(k+1)}$ is the $j$th column cell of the $(k + 1)$ table, $f_{ij}^{(k)}$ is the $j$th column cell of the $k$th iteration table, $f_{+j}^*$ is the $j$th column target total (1/2 in this example), and $f_{+j}^{(k)}$ is the $j$th column total of the $k$th table. After the second iteration ($k = 2$), the column totals are $f_{+j} = (1/2, 1/2)$, which matches the desired target. However, the row totals no longer match their targets. Therefore, the iteration process continues, alternating between the row and column scaling, until both row and column totals converge to the target values of $f_{i+}^* = f_{+j}^* = (1/2, 1/2)$, as shown in the final table of Figure D1 ($k = 6$).

Importantly, the initial table ($k = 0$) and the standardized table ($k = 6$) have approximately the same odds ratio (around 12). This means that that the core pattern of association of the initial table has been preserved during the standardization process, as measured by the odds ratio.

Figure D1: Example of IPF standardization method

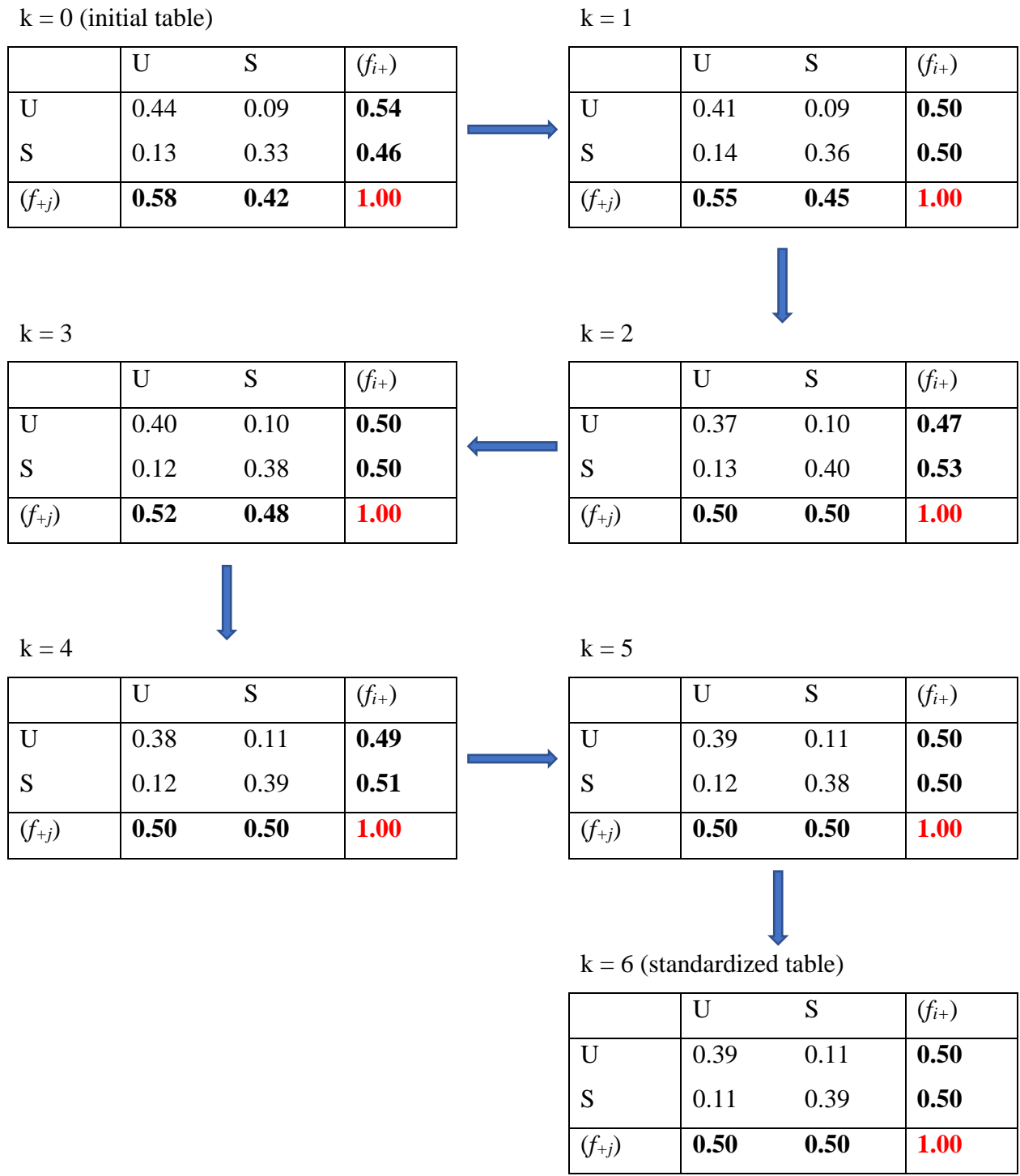